# Multiview Scattering Scanning Imaging Confocal Microscopy through a Multimode Fiber

SAKSHI SINGH,[1, *] SIMON LABOUESSE,[1] AND RAFAEL PIESTUN[1]

*Department of Electrical, Computer and Energy Engineering, University of Colorado Boulder, Colorado, 80309, USA*
*\*sakshi.singh@colorado.edu*

**Abstract:** Confocal and multiphoton microscopy are effective techniques to obtain high-contrast images of 2-D sections within bulk tissue. However, scattering limits their application to depths only up to ~1 millimeter. Multimode fibers make excellent ultrathin endoscopes that can penetrate deep inside the tissue with minimal damage. Here, we present Multiview Scattering Scanning Imaging Confocal (MUSSIC) Microscopy that enables high signal-to-noise ratio (SNR) imaging through a multimode fiber, hence combining the optical sectioning and resolution gain of confocal microscopy with the minimally invasive penetration capability of multimode fibers. The key advance presented here is the high SNR image reconstruction enabled by employing multiple coplanar virtual pinholes to capture multiple perspectives of the object, re-shifting them appropriately and combining them to obtain a high-contrast and high-resolution confocal image. We present the theory for the gain in contrast and resolution in MUSSIC microscopy and validate the concept through experimental results.

## 1. Introduction

Confocal microscopy [1,2] is a widely used technique that enables optical sectioning for imaging with high contrast within scattering tissue. It employs a scanning focal spot to sequentially sample small segments of the object followed by filtering of the backscattered light using a small pinhole in the scanning-spot conjugate plane blocking the out-of-focus light. In practice, the pinhole diameter is chosen to be large enough to achieve a desired tradeoff between optical sectioning/resolution and signal integrity. The technique has been widely successful, enabling for instance, clinical studies for imaging of the cornea [3,4], imaging in body cavities using fiber-optic catheters [5,6] and skin cancer detection [7,8]. However, up to date confocal imaging in the deep tissue regime remains infeasible due to the highly scattering nature of tissue and associated insufficient signal-to-noise ratio (SNR) levels.

Multiphoton microcopy is another effective approach to achieve optical sectioning with improved penetration depth. It provides intrinsic optical sectioning without needing to filter the backscattered light through a pinhole due to the two-photon [9] or multi-photon [10] effect on the optical response. Unlike confocal microscopy, which utilizes only ballistic photons, multiphoton microscopy allows detecting both ballistic and scattered photons [11], hence allowing imaging with better SNR. Furthermore, multiphoton imaging helps achieve penetration depths up to 2 mm using long excitation wavelengths [10,12] or by employing optical clearing [13,14]. However, the penetration depth is still significantly limited.

Alternatively, several endoscopic solutions have been proposed to image deep inside the tissue using single-mode fibers [15–17], fiber bundles [18,19], GRIN lenses [20,21], multicore fibers [22,23], and multimode fibers [24–31]. Among these, multimode fibers (MMFs) make the least invasive and light efficient endoscopes that relay the highest information content in a given cross section. Demonstrations of confocal imaging through multimode fibers have been made by digitally backpropagating from the detector to the object plane and filtering the signal through a virtual pinhole [32,33] or by means of optical correlation [34]. These demonstrations showed imaging of 2-D samples through MMFs with optical sectioning and improved contrast.

However, their application in imaging in thick tissue remains challenging due to SNR limitations.

Confocal microscopy theoretically also has the capability to provide a factor of two in the lateral resolution [35,40] with respect to the diffraction limited resolution based on the Rayleigh criteria [36,37]. However, achieving this gain in resolution is impractical as it requires using a detection pinhole smaller than the size of the scanning focal spot, which compromises the signal strength. Improvement in imaging resolution through multimode fibers has been demonstrated using two-photon imaging [38,39], saturated excitation [40], and by employing a multiple scatterer before the fiber [41,42]. These approaches however come at the cost of expensive short pulse excitation sources, infeasibly high peak power, loss in transmitted light, or need for using short fibers. Another approach used a parabolic tip design [43] to increase the effective NA however the design reduces the field of view and requires a larger working distance, which makes the endoscope susceptible to tissue induced light distortions due to index mismatch. Recently, resolution beyond the diffraction limit [44] has also been demonstrated using MMFs by assuming sparsity in samples [31] however, it requires SNR levels of the sample higher than those feasible with bio-compatible markers.

Here, we present Multiview Scattering Scanning Imaging Confocal (MUSSIC) microscopy through MMFs, an approach to overcome the SNR limitation in confocal microscopy through complex media by employing multiple coplanar virtual pinholes to collect multiple perspectives of the object, followed by proper processing, and combining them to retrieve a high SNR confocal image. Our method builds on the principle of image scanning microscopy (ISM) [35,45–47], transmission matrices [28,48], and digital phase conjugation [24]. ISM is used to boost the SNR in traditional confocal microscopy where the system is shift invariant. However, in contrast with ISM, MUSSIC microscopy does not require a direct measurement of the images of the scanning focal spots. The so-called pixel reassignment operation typical of ISM is performed in a virtual phase-conjugate plane. Moreover, we demonstrate that given the transmission matrix of the system, MUSSIC microscopy can be employed for a more general, shift-variant system such as a generalized complex medium.

We first present a generalized framework to demonstrate the principle of MUSSIC microscopy through complex media and the theory for SNR and resolution gain. Further, we verify the theory experimentally by performing MUSSIC microscopy through an MMF by measuring its transmission matrix (TM). Using the TM, we generate focal spots on the far (distal) end of the MMF. As the focal spots scan the object, we collect the reflected speckle patterns on the MMF's near (proximal) end. Using the MMF's TM, we then back-propagate the collected speckle patterns to the object plane [32,33] to virtually access the scanning focal spot fields and implement MUSSIC microscopy using the generalized pixel reassignment method [45,46]. Our experimental approach is quite general and is also applicable to imaging systems with separate excitation and detection paths [17,49–51]. We evaluate the SNR, optical sectioning and resolution of the reconstructed images and compare our approach with the conventional confocal and single pixel imaging [28,48] approaches.

## 2. Principle of MUSSIC microscopy

Imaging through an MMF is performed by calibrating the relationship between the input and output fields through the system, described by its TM. The TM can be measured experimentally with both phase and amplitude information by sending an orthogonal set of input fields into the system accompanied with a phase-stepping reference field [28,48]. A spatial light modulator (SLM) is typically employed to generate different input fields. Let us denote the different fields propagating through the system by the letter $E$ followed by different superscripts. We assume that the forward TM between the SLM plane to the MMF distal plane is $T$, while the TM from the distal plane to the proximal camera plane is $T^b$.

If the set of fields projected on the SLM are vectorized and stored in the columns of the matrix $E^{in}$, the proximal fields reflected from the MMF are stored in the columns of a matrix

$E^p$, and the object is characterized by a reflection matrix $O$,[1] then the entire system can be described as:

$$E^p = T^b O T E^{in} \quad (1)$$

Let the subscripts denote the row and column indices of the matrices respectively. If we denote the field illuminating the object, as $E^{il} = TE^{in}$, then for the $k^{th}$ incident field $E^{in}_{*k}$, where the asterisk denotes the full set of indices along the particular dimension, the $l^{th}$ pixel of the proximal field, $E^p_{lk}$ is calculated as follows,

$$E^p_{lk} = \sum_{i=1}^{N_{out}} T^b_{li} O_{ii} E^{il}_{ik} \quad (2)$$

Equation 2 shows an overlap function between the excitation and detection point spread functions (PSFs), $T^b_{l*}$ and $E^{il}_{*k}$ weighted by the object reflection function $O$, analogous to the overlap integral used to calculate the resultant field at a confocal pinhole in a conventional confocal microscopy system [36].

Unlike conventional confocal imaging systems which are shift invariant and present a localized PSF, the excitation and detection PSFs for an MMF follow a complex random distribution and are shift variant. Hence, to adopt the raster scan approach for MMF imaging [24,25,28], an input field, $E^{in}_{*k} = T^{\dagger}_{*k}$, must be projected on the SLM to create a diffraction limited focal spot on the $k^{th}$ pixel on the distal end of the MMF. The dagger denotes the conjugate transpose operation. Since the detection path is also through the MMF, the focal spot scanning the object transforms to a speckle pattern on reaching the proximal end of the MMF, hence, in principle, destroying all spatial information.

To reverse the effect of the detection path, we can virtually backpropagate the detected proximal speckle fields to the distal plane [32,33] using the MMF's backward TM, as depicted in the schematic in Fig. 1(a). Interestingly, the above principle is applicable to scattering media in general.

Finally, the virtual distal field denoted as $E^d$, is calculated by taking the product of the proximal fields with the inverse of the backward TM. The TM is however a poorly conditioned matrix, and its inverse does not exist. If we approximate its inverse as its conjugate transpose, as we did earlier for creating phase conjugated focal spots on the distal end, then the backpropagated fields, $E^d$, can be calculated as follows,

$$E^d = (T^b)^{\dagger} E^p = (T^b)^{\dagger} O E^{il} \quad (3)$$

A zoom-in on an example virtual distal field is shown in Fig. 1 (b) where each pixel on the discrete grid serves as a virtual pinhole.

We define $D_{T^b} = (T^b)^{\dagger} T^b$ as the virtual detection PSF of the system. Similarly, we define the pre-SLM to distal plane virtual excitation PSF, $D_T = TT^{\dagger}$, which includes the wavefront projected on the SLM, $T^{\dagger}$, for generating focal spots. .Assuming that a plane wave is incident on the SLM, $D_T$ is also the illumination field matrix, $E^{il}$. The matrices $D_T$ and $D_{T^b}$ have the same strong diagonals similar to a convolution matrix for 2D fields used to represent the TMs of shift invariant systems [52,53]. Their Hadamard product yields the net PSF of the system which is narrower than the individual PSFs as depicted in Figure 1 (c). This narrower net PSF is the source of resolution gain in MUSSIC microscopy, and the resolution enhancement is determined by the size of each virtual pinhole relative to the size of the virtual distal Airy disk. We define 1 Airy unit as the distance from the central peak to the first zero crossing in the virtual distal Airy disks.

Once we obtain the full virtual distal field matrix, $E^d$, the on-axis confocal image is obtained from its main diagonal, $E^d_{kk}$, where $k \in (1, N_{il})$ denotes all distal scan positions. This main diagonal comprises the measurements from the central virtual pinhole, p2, indicated in Fig. 1(b). Similar mutually shifted confocal images are also obtained from the diagonals, $E^d_{lk}$, corresponding to the neighboring pixels of $k$ (such as p1 and p3 shown in Fig. 1 (b)) where

---

[1] Note that if the object is a 2D reflective object, then $O$ is a diagonal matrix.

$l$ takes $N^2 - 1$ values other than $k$ in the $N \times N$- pixel neighborhood of each scan position $k$. All the $N^2$ confocal images can then be re-shifted to a common axis, weighted, and combined to yield a high-SNR MUSSIC image reconstruction as illustrated in Fig. 1(d). We found that a hybrid weighting approach improves reconstruction. In particular, weighting the images obtained from the pinholes within 1 Airy unit by unity and the images obtained from pinholes outside one Airy unit in proportion to their mean value worked best.

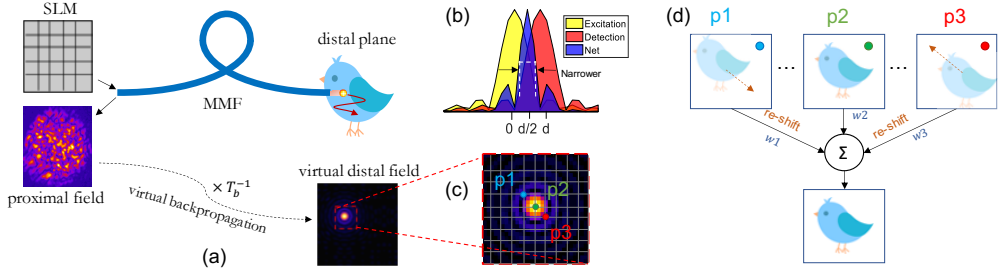

Fig. 1. Principle of MUSSIC microscopy. (a) Illustration of the principle of MUSSIC microscopy through a multimode fiber. An SLM projects the phase patterns to generate scanning focal spots on the distal end, where the object is located. Light reflected from the object couples back into the fiber and reaches the proximal end as a speckle field. The proximal speckle field is recorded and virtually backpropagated to the distal end using the backward TM. This virtual distal field matrix comprises the MUSSIC raw data. (b) **Resolution improvement in MUSSIC microscopy.** The virtual excitation PSF, $D_T$, the virtual detection PSF, $D_{T^b}$, and the net PSF, calculated as the product of the former two are shown in yellow, red, and blue respectively for an example confocal image obtained from a virtual pinhole p$_i$ at a distance $d$ from the on-axis pinhole. Although the net PSF is shifted from the axis by a distance, $d/2$, it is narrower than the former two PSFs, hence leading to an improved resolution. The excitation and detection wavelengths are assumed to be identical here. (c) Zoom-in on the focal spot in the virtual distal field shown in 1 (a), as demarcated by the red dotted line. Each pixel in the field acts as a virtual pinhole. While the central pixel, p2 measures the on-axis confocal image, the pinholes p1 and p3 also measure similar confocal images from different perspectives. (d) **SNR improvement in MUSSIC microscopy.** Illustration of the pixel reassignment algorithm to combine the confocal images obtained from different virtual pinholes. Confocal images from the pinholes p1, p2 and p3, as labelled in 1(c), are shown on the top. Besides the image obtained from the on-axis pinhole, p2, all confocal images are shifted off the optical axis by a distance determined by the location of their corresponding virtual pinholes. By applying appropriate shifts and weights to them and combining them together, a single high-contrast MUSSIC image is obtained.

## 3. Methods

In what follows we present a specific experimental implementation of MUSSIC.

### 3.1. Calibration of forward TM, T

The forward TM $T$ is measured with both phase and amplitude information, by sending orthogonal input fields into the fiber accompanied by a phase-stepping reference field. We choose the plane-wave basis that transforms to focal spots in the Fourier plane, which are then coupled into the MMF. These patterns are constant in amplitude and their phases are modulated using a spatial light modulator (SLM). The SLM's active area is divided into two sections each for a changing grating pattern and a phase-stepping reference frame that surrounds it. The intensity measurements at the fiber output for each projected pattern, as the reference field is phase stepped, allow for the recovery of the output fields [28,48]. These output fields are vectorized and used to build all the rows of the matrix $T$.

### 3.2. Calibration of backward TM, $T^b$

The TM of an MMF obeys the reciprocity rule. However, in practice, we measure the TM between the SLM and the distal plane and the reciprocity assumption only holds true if the detection plane perfectly matches the SLM plane in scale and orientation. This is challenging in practice, requiring a sensitive and time-consuming alignment procedure [26]. Moreover, oftentimes, it is desirable to separate the collection and detection pathways in endoscopes to improve throughput or to gain some feedback from the distal end [17,49–51]. In such cases $T^b \neq T'$. For other modalities like fluorescence imaging, the excitation and detection PSFs are different by default due to difference in the excitation and fluorescence wavelengths. With these considerations, here we propose a separate calibration of the matrix $T^b$ from the distal plane to the detector plane.

Towards this end, we place a mirror at the distal end of the fiber and scan focal spots on it, while measuring the reflected fields on the proximal end, denoted as $E^{p-mirror}$. These measurements give us an estimate of $T^b$, which we denote as $T^b_{obs}$, as follows

$$T^b_{obs} = E^{p-mirror} = T^b \, I \, E^{il} \qquad (4)$$

The matrix $I$ in the above equation represents the mirror reflection matrix, which we assume to be an identity matrix. The distal fields are then given by

$$E^d_{obs} = \left(T^b_{obs}\right)^{\dagger} E^p = (E^{il})^{\dagger} D_{T^b} \, O \, E^{il} \qquad (5)$$

In comparison to Eq. 3, the above equation has an additional term $(E^{il})^{\dagger}$ on the right-hand side because of our double pass approach for calibration of $T^b$. Since we use a raster scan approach, both the $E^{il}$ and $(E^{il})^{\dagger}$ matrices have the same prominent diagonals as a convolution matrix with a diffraction limited Gaussian kernel and the distal fields obtained from Eq. 5 are a good approximation of the distal fields calculated in Eq. 3, hence enabling confocal and MUSSIC image reconstruction. Moreover, the theoretical resolution gain of confocal imaging is also preserved as the bandwidth of the terms on the left and right of the object, O, in the above equation remain unchanged.

### 3.3. Optimal inversion of backward TM

As mentioned earlier, we can use the conjugate transpose operator when the inverse of a matrix does not exist. This method works well for generating phase conjugated focal spots, as required when raster scanning on the distal side of the fiber. However, when calculating the backpropagated distal fields, the conjugate transpose is not the best inversion method. We can optimize the inversion of the backward TM using a Tikhonov regularization technique [32,54]. This involves computing the singular value decomposition of the backward TM, $T^b_{obs} = USV^{\dagger}$ and finding its inverse as follows

$$T^b_{obs}{}^{RI} = VS^{RI}U^{\dagger} \qquad (6)$$

$S^{RI}$ is the regularized inverse of the diagonal matrix of singular values, $S$, calculated by replacing the singular values $\sigma_i$ in the diagonal of S with $\frac{\sigma_i}{\sigma_i^2+\beta^2}$, where $\beta$ is the regularization parameter. In experiments, we find that by calculating the backpropagated distal fields using Tikhonov regularized inverse of the $T^b_{obs}$, instead of $\left(T^b_{obs}\right)^{\dagger}$ in equation 5, yields image reconstructions with improved SNR and contrast. For our results, we chose a $\beta$ value equal to 10% of the highest singular value of the backward TM. A comparison of reconstructions using the conjugate transpose and Tikhonov regularization reveals that although the regularization considerably improves the image quality, the faster reconstruction obtained from the conjugate transpose of the TM also provides a good estimate of the object.

### 3.4. Lowpass filtering and normalization

We perform digital lowpass filtering to bandlimit the spatial frequencies in the acquired data. This eliminates the noise in the high frequencies and ensures that all acquired images have speckles with a minimum grain size limited by diffraction. The frequency cutoff is chosen by computing the average of the Fourier transform of all the measured proximal images and setting the values below a minimum threshold outside the central Gaussian peak to zero.

The MUSSIC reconstruction of a blank object or a perfectly reflective mirror is a speckle pattern. This is explained by the fact that although the virtual detection PSF share the same prominent diagonals and are similar in structure to a convolution matrix, they also have non-zero values outside of those diagonals. This is also true for the virtual excitation PSF and explains the intensity variations in the focal spots used to scan the object. In order to account for this non-uniformity, we normalize the reconstructed confocal and MUSSIC reconstructions w.r.t to their "blank" counterparts i.e., the reconstruction images obtained when a mirror is placed at the distal end. This normalization significantly improves the image quality.

*3.5. Imaging without full field backpropagation*

Calculating the full matrix $E^d$, involves heavy computation, with a complexity O ($N_{il}^2 N_{in}$). However, in fact, access to the full backpropagated distal fields is not necessary to calculate the confocal or MUSSIC images. The only data points required in each distal field are in the neighborhood of the scanning focal spot, for every scan position. This number, which we define as $N_{pinholes}$ is chosen to be roughly equal to the number of pixels that sample a focal spot and is much lower than the number of illumination patterns used for imaging. Hence, if we compute only the desired diagonals from the matrix $E^d$ corresponding to the $N_{pinholes}$ neighboring pixels, the complexity of the calculation drops down to only O ($N_{pinholes} N_{in} N_{il}$) for the MUSSIC image and to only O ($N_{in} N_{il}$) for a single confocal image. When using the conjugate transpose of the backward TM to invert it, this method for obtaining the confocal image is similar to the correlation method [32]. Imaging with this reduced computation approach, which we call fast-MUSSIC, enables MUSSIC reconstruction of a 20,000-pixel image in 4 minutes on a DELL Desktop computer with a 3.2 GHz Intel Core i5 processor and 64 GB RAM.

## 4. Experimental Setup

The experimental setup used to demonstrate MUSSIC microscopy through an MMF is illustrated in figure 2. We use a 785 nm CW Crystal laser and a Meadowlark optics liquid crystal SLM (HSPDM 512) for phase modulation. The laser beam goes through a half waveplate and polarizer for polarization control, followed by a 4-F system to match the beam diameter to the active area of the SLM. The SLM plane is then imaged onto the back-aperture of a microscope objective, OBJ 1, which couples the light into the MMF. We used a step-index fiber of diameter 50 $\mu$m and 0.22 numerical aperture (NA) for all our experiments. A polarizing beam-splitter (PBS) between the SLM and OBJ 1 is used to direct the back-reflected light from the fiber onto a camera, CAM 2. A half waveplate before the PBS allows controlling the polarization axis of the incident beam and a quarter waveplate along with the PBS act as an optical isolator to prevent back-reflections from the proximal facet of the fiber from reaching the camera. The distal facet of the MMF is imaged onto a camera, CAM 1 using another lens during the forward TM calibration. A polarizer before the camera allows detection of only one polarization component.

After the forward TM calibration, a mirror is placed near the fiber distal tip for calibration of the backward TM. The backward TM is calibrated using back-reflected fields on the proximal side of the fiber, while focal spots are projected on the distal side. A phase shifting reference frame is simultaneously projected on the SLM along with the phase conjugated patterns for distal raster scan, for measuring both the phase and amplitude of the back-reflected fields. The back-reflected light from the mirror couples back into the fiber and is detected on the proximal side using another camera, CAM 2. This camera images the back-aperture of the microscope objective OBJ 1 using another 4-F system and is placed in a plane equivalent to the SLM plane. A polarizer before the camera allows detection of a single polarization component.

After both calibrations, the sample to be imaged replaces the mirror at the distal facet of the MMF, and the back-reflected fields from the object are recorded as it is raster scanned.

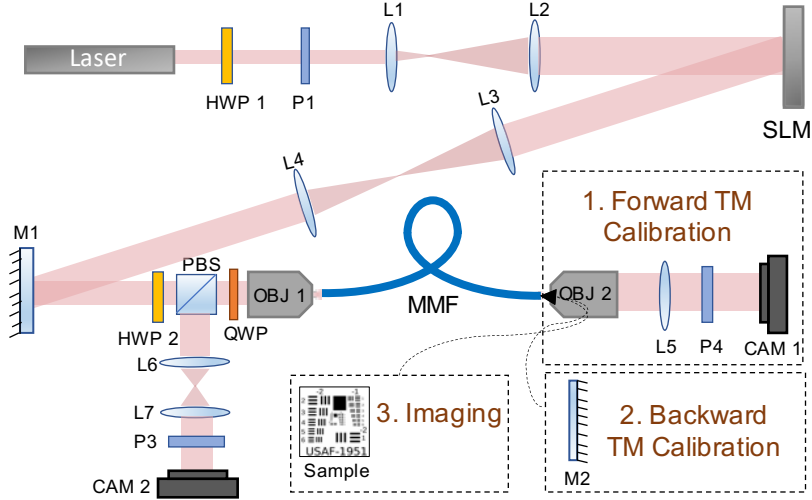

Fig. 2. Experimental setup for MUSSIC microscopy through an MMF. HWP 1-2: Half waveplates, P1-3: linear polarizers, L1-7: lenses, MMF: 10 cm long step-index fiber from Thorlabs (FG050LGA), PBS: polarizing beam-splitter, M1-2: mirrors, QWP: quarter waveplate, OBJ 1-2: microscope objectives, CAM 1-2: cameras to measure distal and proximal intensities.

## 5. Results

### 5.1. SNR and resolution analysis

We perform confocal and MUSSIC microscopy in simulation and compare the SNR and resolution of the reconstructed images. We model the MMF TM as a complex random matrix and reconstruct the image of a quadrant of the binary siemens star using the simulated proximal speckle fields, following the backpropagation process described earlier. We added Gaussian noise with 5% variance to the simulated proximal fields before the image reconstruction. Each virtual pinhole in our simulation has a radius of 0.11 Airy unit (a.u.). Hence one Airy disk spans across 9 x 9 individual pinholes.

The ground truth object and its confocal and MUSSIC reconstructions are shown in Figs. 3 (a-d). Figs. 3 (b,c) show the confocal reconstructions using a 3 x 3 macro-pinhole and a 9 x 9 macro-pinhole respectively. Fig. 3 (d) uses the same group of 9 x 9 pinholes as 3 (c) but employs the MUSSIC approach. We find that although the SNR improves significantly among the confocal image reconstructions as the size of the macro-pinhole increases, the resolution degrades. On the other hand, the MUSSIC reconstruction, which uses the same group of pinholes as the second confocal reconstruction retains the high-SNR, while also preserving the resolution. The difference in resolution can be more clearly visualized in Fig. 3 (e) that shows the normalized cross sections in the image reconstructions corresponding to the green solid lines in Figs. 3 (a-d). We find that confocal reconstruction with the 1 a.u. pinhole fails to resolve the image features, while the MUSSIC reconstruction using the same raw data resolves them just as well as the confocal reconstruction with the small 0.33 a.u. pinhole.

Next, we analyze in Fig. 3 (f) the reconstruction error and correlation as a function of the number of pinholes used, for the green line cross-sections marked in Fig. 3 (a-d) in the absence of noise. We find that the cross-section error and correlation w.r.t. the ground truth increases and decreases respectively as the number of pinholes constituting the macro-pinhole increases for the confocal reconstruction. On the other hand, both metrics for the MUSSIC reconstructions remain unaffected, indicating that the image quality is preserved.

Furthermore, in Fig. 3(g), we compare the average frequency response for the confocal and MUSSIC methods obtained for a point object using 81 pinholes. These responses were also obtained in the absence of noise. We find that the frequency cutoff of the MUSSIC reconstruction is almost double that of the OTF of the system, which is the theoretically claimed gain in resolution according to Rayleigh's criterium [36,37]. On the other hand, the confocal reconstruction obtained from the 1 a.u. pinhole has a frequency cutoff 1.4 times higher than that of the system OTF. Next, we computed the noiseless PSFs for the confocal and MUSSIC methods and plotted the full width half maxima (FWHM) of the PSFs as a function of the number of pinholes used. We find that the FWHM for the confocal reconstruction increases with the number of pinholes constituting a macro-pinhole, while the FWHM for the MUSSIC reconstruction remains unchanged [Fig. 3(h)].

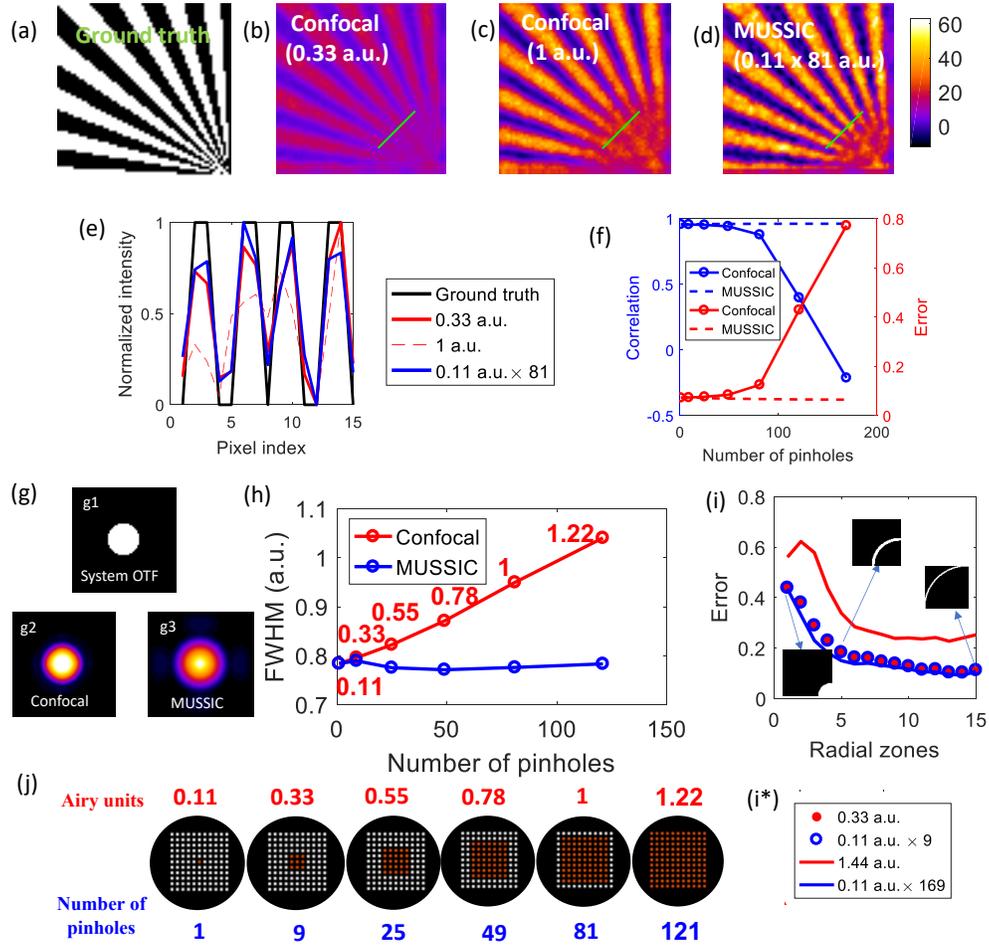

Fig. 3. Comparison of confocal and MUSSIC image reconstructions. a) Binary object (ground truth), (b) confocal image obtained from a (b) 3 x 3 macro-pinhole of size 0.33 a.u. and (c) a 9 x 9 macro-pinhole of size 1 a.u. (d) MUSSIC image obtained using 81 pinholes, each of radius 0.11 a.u. (e) Plot of the normalized cross sections indicated by the solid green lines in (a-d). f) Error and correlation of confocal and MUSSIC images as a function of the number of pinholes used, as measured for the cross section marked by the green solid lines shown in (a-d). (g) Optical transfer function of the system (g1) and the average frequency response of the system for a point object obtained by performing confocal (g2) and MUSSIC microscopy (g3) using 81 pinholes. (h) Full width half maxima of the net PSFs obtained from confocal and MUSSIC reconstructions as a function of the number of pinholes used. The net pinhole size in Airy units is indicated at various points in red font for the confocal curve. (i) Root mean square error for 15 annular regions of increasing radii starting from the bottom right, for normalized confocal images (blue) and MUSSIC images (red) using 9 and 169 pinholes. The insets illustrate the annular regions 1, 5 and 15 respectively. (j) Schematic of the various pinholes used in confocal and MUSSIC reconstructions. All pinholes are shown in white, while the pinholes used for reconstruction are shown in red. Their sizes in Airy units and number of pinholes are also indicated in red and blue respectively.

Finally, in Fig. 3 (i), we compare the root mean square error of different normalized reconstructions as a function of the annular radius measured from the center of the Seimens star on the bottom right corner of Fig. 3(a). For this comparison, no noise was added to the reconstructions to analyze the effect of using increasing number of pinholes on resolution. We divide the image quadrant into 15 radial zones and plot the error w.r.t. the ground truth image in each zone for the different reconstruction methods. We find that while the error for the confocal reconstruction images increases with the radius of the macro-pinhole, the error in the MUSSIC reconstruction remains almost unchanged as the number of used pinholes increases from 3 x 3 to 13 x 13. The inset images in the figure show the radial zones 1, 5 and 15 from left to right. Fig. 3 (j) shows a schematic of the different pinhole groups used for confocal and MUSSIC reconstructions along with their respective sizes in Airy unit and number of pinholes.

### 4.3. Experimental results

We demonstrate MUSSIC microscopy through a multimode fiber and compare the reconstructions in Fig. 4. The field of view (FOV) consists of the fourth and fifth elements of the 7$^{th}$ group in the USAF 1951 resolution target, which have a resolution of 181- and 203-line pairs/mm respectively. Figs. 4 (a-c) show the MUSSIC reconstructions as the number of used pinholes increases from 1 to 81. We see that the reconstruction SNR improves consistently with the number of pinholes.

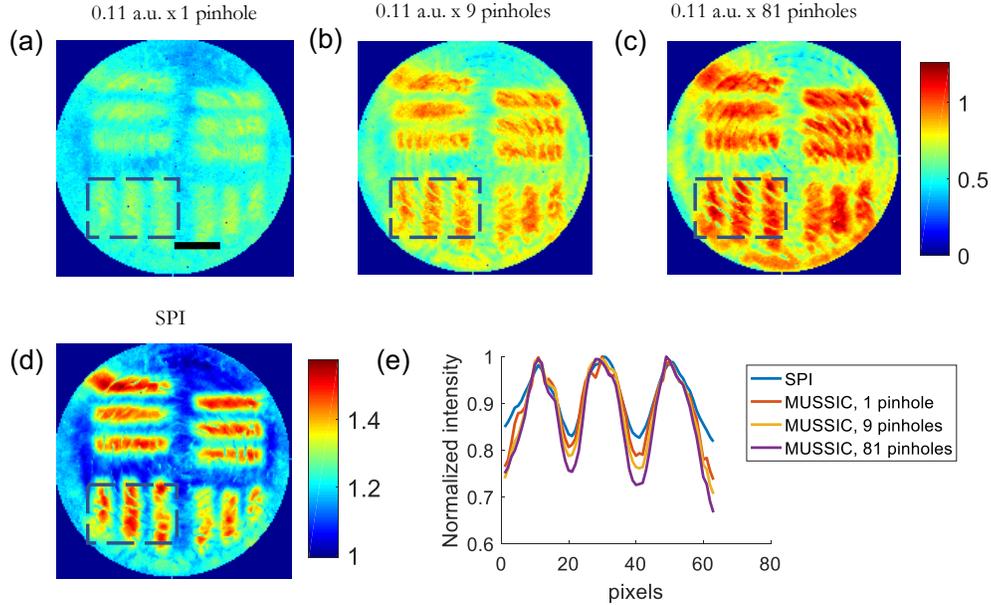

Fig. 4. Demonstration of MUSSIC microscopy through a multimode fiber. The object is a binary USAF 1951 resolution target. The FOV is a 50 microns wide-160 x 160-pixel window. Scale bar (black) is 10 $\mu$m. (a-c) Comparison of MUSSIC images obtained using (a) 1, (b) 9 and (c) 81 pinholes respectively. (d) SPI image obtained by integrating the absolute value of the proximal fields. (e) Average cross section along the horizontal direction obtained from the cropped regions indicated by dashed lines in the images in (a-d). Postprocessing for the MUSSIC images involved regularized TM inversion and normalization, as explained in the methods section.

Fig. 4(d) shows the single pixel image (SPI) [28], obtained by integrating the absolute values of all the pixels in the proximal speckle fields when the distal plane is raster scanned. Fig. 4 (e) shows the normalized average cross sections for all the reconstructions along the horizontal direction for a cropped window within the FOV (grey-dashed lines). The SPI image, obtained without virtual backpropagation and using the signal from all the distal pinholes, has a higher SNR than other reconstruction but lower contrast. For the MUSSIC reconstructions the contrast improves with the number of pinholes.

Next, we demonstrate the optical sectioning capability of the MUSSIC technique. The FOV shows the first element of the 7$^{th}$ group in the resolution target. We move the 2D target in steps of 20 $\mu$m in the axial direction and away from the fiber distal facet and capture the back-reflected fields from the object at three z-positions. We compare the SPI reconstructions with the MUSSIC reconstructions at the three positions in Figure 5. We observe that the object almost disappears in the background already after a z-shift of 20 $\mu$m in the case of the MUSSIC images, while the SPI reconstructions carry a significant amount of light from the sample even after a z- displacement of 40 $\mu$m. Hence the MUSSIC approach performs better in rejecting the background of the image.

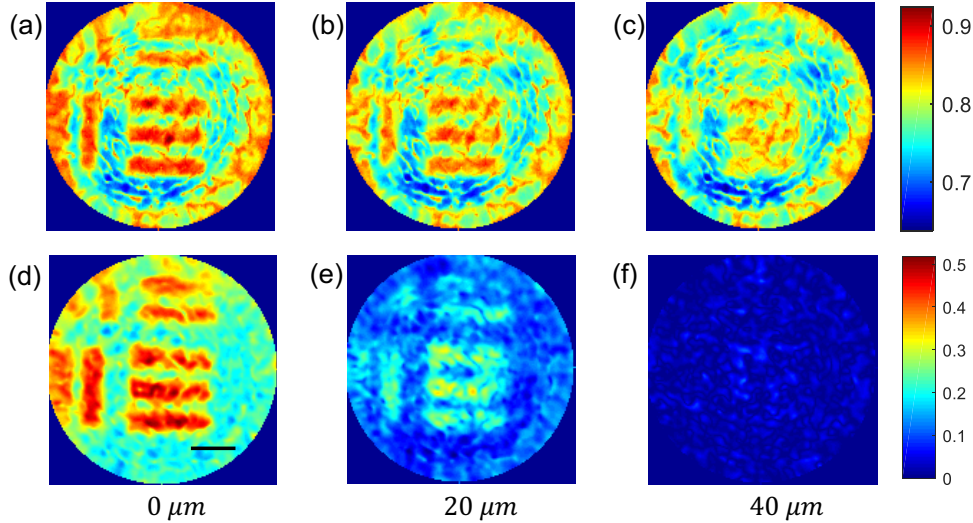

Fig. 5. Comparison of optical sectioning in MUSSIC microscopy and SPI through multimode fibers. (a, b, c) SPI reconstruction at z=0, 20 and 40 $\mu$m respectively. (d, e, f) MUSSIC reconstruction from 25 pinholes at z=0, 20 and 40 $\mu$m respectively. Postprocessing for the MUSSIC images involved regularized TM inversion, lowpass filtering and normalization, as explained in the methods section. Scale bar is 10 $\mu$m.

## 6. Discussion and conclusions

We have demonstrated MUSSIC microscopy through a multimode fiber to enable imaging with improved optical sectioning, high SNR, and improved resolution. By increasing the number of virtual pinholes used for reconstruction, the image contrast improves with respect to the virtual confocal image. The tradeoff is in the computational complexity which grows linearly with the number of pinholes.

In the experiments, as usual, the quality of the reconstruction and in particular the achieved resolution is affected by overall SNR. While we demonstrate a factor of 2 improvement in the spatial frequency bandwidth of MUSSIC images relative to the optical transfer function of the system in simulation [Fig 3 (g)], a similar improvement is difficult to observe or characterize in the noisy experimental images. This can be attributed to several uncontrolled experimental factors. Firstly, the image quality is dependent on the accuracy of the reconstructed virtual distal fields, which is in turn determined by the quality of the inverse estimate of the backward TM. Secondly, our technique assumes that the object is illuminated with focal spots with no background speckle. In practice, the enhancement of the focal spots, defined as the ratio of focal spot intensity and the background speckle intensity, is limited and the background speckle contributes to noise in the reconstruction. Moreover, in the implemented experimental calibration of the backward TM, we assumed a perfect reflective mirror whose reflection matrix is an identity matrix. In practice, some light is lost at the mirror and does not couple back into the fiber. Moreover, the object must be positioned precisely in the plane of the mirror used during the calibration of the backward TM. Any deviation would cause the signal from the object plane to become out-of-focus which would consequently be rejected by the virtual pinholes.

Furthermore, to keep our experimental setup simple and robust to thermal and mechanical fluctuations, we used an internal reference for phase measurements which transforms to a non-uniform speckle in the plane of interest with many intensity nulls, also known as blind spots. The field from these blind spots cannot be recovered, which further degrades the image reconstruction quality. Using complementary reference speckles [55,56] or an external plane wave reference are possible ways to eliminate the blind spots, although they either require increased measurement time or a more complex setup with phase tracking to account for phase drifts. Bending sensitivity of the fiber is another challenge and any perturbations after calibration lead to systematic errors in the image reconstruction. However, various approaches exist to mitigate fiber perturbations, including feedback mechanisms [51,57–61], fiber selection [62] or illumination pattern optimization [63,64].

Here we limit our experiments to the coherent imaging modality, but the high SNR capability of MUSSIC microscopy paves a feasible path to fluorescence imaging. Calibration of the multispectral TM of scattering media has been demonstrated in multiple reports [65–67]. With the help of the multispectral TM, one could for instance, scan multi-spectral focal spots on the proximal side while speckle patterns are projected on the object at the distal end. With knowledge of the distal intensity patterns, the object can be recovered [31,33]. An advantage of scanning focal spots on the proximal side is that it would eliminate the need for coherent backpropagation and enable imaging by solving a simpler intensity-only inverse problem.

A further generalization of the technique can be made by choosing distal illuminations that are not focal spots, but arbitrary speckle patterns. Speckle illumination is ideal for compressive sampling and can enable imaging with fewer illumination patterns and shorter data acquisition times [31]. Furthermore, it can also eliminate the need for wavefront shaping.

In summary, this report demonstrates the capability of MUSSIC microscopy in enabling high SNR and high-resolution imaging through a MMF endoscope. Given the generalized principle of the technique, its application is not limited to the raster scan approach or to MMF and can easily be adapted to other endoscopic probes that might require different excitation and detection paths such as double-clad fibers.

## Acknowledgements

We thank Dr. Antonio M. Caravaca for several fruitful discussions that helped improve this manuscript. We thankfully acknowledge funding support from the National Science Foundation (award 1611513), the Colorado Office of Economic Development (Award DO 2020-2464) and the University of Colorado through the Lab Venture Challenge.## References

1. R. H. Webb, "Confocal optical microscopy," Rep. Prog. Phys. **59**, 427–471 (1996).
2. S. W. Paddock, "Principles and practices of laser scanning confocal microscopy," Mol. Biotechnol. **16**, 127–149 (2000).
3. I. Jalbert, F. Stapleton, E. Papas, D. F. Sweeney, and M. Coroneo, "In vivo confocal microscopy of the human cornea," Br. J. Ophthalmol. **87**, 225–236 (2003).
4. W. Wiegand, A. A. Thaer, P. Kroll, O.-C. Geyer, and A. J. Garcia, "Optical Sectioning of the Cornea with a New Confocal In Vivo Slit-scanning Videomicroscope," Ophthalmology **102**, 568–575 (1995).
5. R. Kiesslich, M. Goetz, J. Burg, M. Stolte, E. Siegel, M. J. Maeurer, S. Thomas, D. Strand, P. R. Galle, and M. F. Neurath, "Diagnosing Helicobacter pylori In Vivo by Confocal Laser Endoscopy," Gastroenterology **128**, 2119–2123 (2005).
6. P.-L. Hsiung, J. Hardy, S. Friedland, R. Soetikno, C. B. Du, A. P. Wu, P. Sahbaie, J. M. Crawford, A. W. Lowe, C. H. Contag, and T. D. Wang, "Detection of colonic dysplasia in vivo using a targeted heptapeptide and confocal microendoscopy," Nat. Med. **14**, 454–458 (2008).
7. M. Rajadhyaksha, M. Grossman, D. Esterowitz, R. H. Webb, and R. Rox Anderson, "In Vivo Confocal Scanning Laser Microscopy of Human Skin: Melanin Provides Strong Contrast," J. Invest. Dermatol. **104**, 946–952 (1995).